\documentclass[12pt]{article}
\usepackage{geometry}
\usepackage{xcolor}
\usepackage{a4}
\usepackage{graphicx}
\usepackage{epsf}
\usepackage{amsmath}
\usepackage{amssymb}
\usepackage{cite}
\newcommand{\be}{\begin{equation}}
\newcommand{\ee}{\end{equation}}

\newcommand{\Rmnum}[1]{\expandafter\@slowromancap\romannumeral #1@}
\newcommand{\bea}{\begin{eqnarray}}
\newcommand{\eea}{\end{eqnarray}}
\begin{document}
\def\C{{\mathbb{C}}}
\def\R{{\mathbb{R}}}
\def\s{{\mathbb{S}}}
\def\T{{\mathbb{T}}}
\def\Z{{\mathbb{Z}}}
\def\W{{\mathbb{W}}}
\def\Bbb{\mathbb}
\def\BZ{\Bbb Z} \def\BR{\Bbb R}
\def\BW{\Bbb W}
\def\BM{\Bbb M}
\def\BC{\Bbb C} \def\BP{\Bbb P}
\def\CP{\BC\BP}
\begin{titlepage}
\title{Fermi Normal Coordinates and Fermion Curvature Couplings in General Relativity} \author{} 
\date{
Anshuman Dey, Abhisek Samanta, Tapobrata Sarkar 
\thanks{\noindent 
E-mail:~ deyanshu, rupam, tapo @iitk.ac.in} 
\vskip0.4cm 
{\sl Department of Physics, \\ 
Indian Institute of Technology,\\ 
Kanpur 208016, \\ 
India}} 
\maketitle 
 
\abstract{We study gravitational curvature effects in circular and radial geodesics in static, spherically symmetric space-times, using
Fermi normal coordinates. We first set up these coordinates in the general case, and then use this to study effective magnetic fields due to gravitational curvature in 
the exterior and interior Schwarzschild, Janis-Newman-Winicour, and Bertrand space-times. We show that these fields can be large for specific parameter values in
the theories, and thus might have observational significance. We discuss the qualitative differences of the magnetic field for vacuum space-times and for those 
seeded by matter. We estimate the magnitude of these fields in realistic galactic scenarios and discuss
their possible experimental relevance. Gravitational curvature corrections to the Hydrogen atom spectrum for these space-times are also discussed briefly.}
\end{titlepage}

\section{Introduction}
\label{intro}
Einstein's General Theory of Relativity (GR) \cite{Weinberg} is an established theory of gravity, and many experimental tests of this theory are well known. 
In particular, studying the effect of space-time curvature on spin systems has a long history (see, e.g \cite{mash1}) starting from the celebrated work of de Sitter almost a century back
\cite{ds}. Indeed, recent experimental results from the Gravity Probe B experiment \cite{probeB} have successfully demonstrated the geodetic and frame-dragging effects, further
strengthening the basis of GR.

Another interesting direction of work has been the analysis of gravitational effects on elementary particles. For example, the Dirac equation was analysed in general curved space-times
by Parker \cite{parker1}, \cite{parker2}, \cite{parker3}, who calculated the shifts in the Hydrogen atom spectra due to gravity. His results showed that for example, at the 
surface of the Sun, the typical contribution of gravity effects will result in shifting the energy levels of the Hydrogen atom to the order of $10^{-50}~{\rm GeV}$, which is
unfortunately too small to detect by present day experiments. Along these lines, one can also think of possible violations of Lorentz and CPT invariance due to gravitational effects (see, e.g \cite{kostelecky1}). 
The EOT-Wash experiment \cite{Eotwash} has been actively seeking to detect such effects via torsion balance measurements, and one of its objectives is to measure gravitational interactions that couple 
to the spin of elementary particles. Here, the effect of gravity on an individual spin is magnified by considering a balance consisting of a very large number of such spins, and bounds
have been placed on magnetic fields that can arise due to gravitational effects. 

Although a preliminary analysis of gravitational effects on particles show that these might be hard to detect with present day experiments \cite{mohanty}, it is important and interesting
to understand scenarios where these might be large. In fact, one might hope to understand qualitative features of space-time itself, by analysing such large effects, if
they exist. Naturally, gravitational effects should maximize near large gravitating bodies, and GR predictions of physical effects in this regime might be of interest in futuristic experiments. 
For example, a concrete question to ask is whether there is any observable enhancement of gravitational effects on spinors near a massive object like a black hole or a neutron star. Or, we
could try to determine such effects near a galactic centre which is dominated by dark matter. There are two immediate problems that arise here. Firstly, any such computation will be 
model dependent, and more importantly, a general relativistic prediction of a physical effect is observer dependent, and one needs to take care of subtleties regarding the latter. There
is ample literature on the subject (see, e.g \cite{bani}) which deals with important phenomenological issues, and in this work we try to complement these by a more formal approach. 

Consider an observer in free fall, i.e in geodesic motion in a gravitational background. We assume that the motion does not back react on the metric, i.e our observer is treated as a 
test particle in a gravitational background. Now say our observer does an experiment on a spin system. 
By writing the Dirac Lagrangian appropriate to a curved background, one can show that this will have fermionic 
pseudo vector couplings, which in the non-relativistic limit reduce to an interaction energy of the form ${\vec s}.{\vec b}$ with ${\vec s}$ being the spin of the particle \cite{kostelecky1}, 
\cite{kostelecky2}, \cite{mash2}. This is what our geodesic observer seeks to determine, perhaps with a torsion balance setup. 
To calculate the effective magnetic field ${\vec b}$ due to gravitational interactions (to be distinguished from the intrinsic magnetic field that arise due to 
motion of charged particles), we need to make a choice of the coordinate system. A natural choice, motivated from 
physical considerations is the Fermi normal coordinate system, first introduced by Manasse and Misner \cite{ferminormal}. 
In Fermi normal coordinates, the metric is locally flat all along the geodesic, i.e the Christoffel connections vanish everywhere on the geodesic, although its
derivatives may not be zero. In this coordinate frame, our locally flat observer can measure an effective magnetic field acting on the spin of the particle. The advantage of using 
Fermi normal coordinates is that one can talk about measurements carried out close to a singularity and not necessarily in a weak field limit. We can therefore address the question of the effect of 
gravity when it is large (of course possible quantum gravity corrections are ignored). Although the implications of such measurements in present day experiments is far from obvious, 
it might encode valuable insights into the nature of space-time.  

In this paper, we carry out such an analysis of observers in circular and radial geodesic motion for the Schwarzschild, Janis-Newman-Winicour (JNW) and Bertrand space-times. While the former is a 
vacuum solution of gravity, JNW space-times are sourced by a scalar field that satisfies the Einstein-Klein-Gordon equations and the Bertrand space-times (BSTs) are seeded by matter
that be given an effective two-fluid description \cite{dbs1}, \cite{dbs2} and can be thought of as a candidate for galactic dark matter. We highlight the important differences between the
nature of gravitational couplings on fermions in these space-times, and show that for Schwarzschild, JNW and BST backgrounds, such couplings can be large on highly relativistic
orbits. For BST backgrounds, our results indicate that it might be possible to obtain indications of galactic dark matter in futuristic experiments.

This paper is organized as follows. In the next section, after briefly reviewing aspects of general spherically symmetric metrics in Fermi normal coordinates, 
we analyse observers in circular geodesic motion and calculate
the effective magnetic field due to curvature couplings in Schwarzschild, JNW and BSTs. In section 3, we follow up this analysis for radial motion in the
same backgrounds. We comment on the important differences in the results. Section 4 ends the paper with a summary and possible directions for future research, where we also 
present some results on the issue of gravitational corrections to the Hydrogen atom spectrum in BST and JNW backgrounds. For the sake of completeness, we list, in two appendices,
the nonzero components of the Riemann curvature tensor in Fermi normal coordinates for circular geodesics in Schwarzschild and Bertrand space-times. 

\section{Circular Geodesics and Fermion Curvature Couplings}

We now consider an observer in circular geodesic motion and set up Fermi normal coordinates to compute fermion curvature couplings. We first set up the general formalism to be 
used in this section and later on in the paper. We begin with some statements which will set the notation and conventions, and will motivate the rest of the work. 

\subsection{Fermi Normal Coordinates and Effective Magnetic Fields}

In GR, in order to connect any result to possible experiments, we need a coordinate system in which the metric is locally flat along the entire geodesic on which our observer moves.
As alluded to in the introduction, such a system was envisaged in \cite{ferminormal} by Manasse and Misner, and let us briefly recapitulate their construction,
which involves a number of steps. First,
we choose an arbitrary point on the geodesic $G$ as the origin, and set up a tetrad basis there. This basis is now parallely transported along the geodesic $G$ and for any event at a point $P$ 
with Fermi normal coordinates $x^{\alpha}$, the time $x^0$ is the proper time along $G$ at the intersection of a space-like hypersurface containing $P$, with $G$. 
The other components of $P$ are obtained by proceeding along a space-like geodesic on the hypersurface from the point of intersection. The tetrad basis (called the Fermi normal basis)
should satisfy \cite{ferminormal}
\begin{equation}
{\hat e}_{\alpha}.{\hat e}_{\beta} = \eta_{\alpha\beta},~~~\nabla_{\nu'}\left({\hat e}_{\alpha}^{\mu'}\right){\hat e}_0^{\nu'}=0,
\label{tetradmain}
\end{equation}
where $\nabla$ is the covariant derivative. Here, primes denote coordinates in which the original space-time metric is written (i.e Schwarzschild or JNW or 
BST coordinates) and the unprimed indices will denote Fermi normal coordinates. Once we have set up the tetrad basis according to the above prescription, we can write down 
the components of the curvature tensor from those of the original space-time, using the explicit forms for the tetrads. These are given by
\begin{equation}
R_{\alpha\beta\gamma\delta} = {\hat e}^{\mu'}_{\alpha}{\hat e}^{\nu'}_{\beta}{\hat e}^{\rho'}_{\gamma}{\hat e}^{\sigma'}_{\delta}R_{\mu'\nu'\rho'\sigma'}
\label{Rfr}
\end{equation}
Having obtained these, the metric around a geodesic $G$, to second order in the coordinates can be shown to be given by \cite{ferminormal}
\begin{equation}
g_{00} = -1 + R_{0l0m}x^lx^m|_G,~~~g_{0i} = \frac{2}{3}R_{0lim}|_Gx^lx^m,~~~g_{ij} = \delta_{ij} + \frac{1}{3}R_{iljm}|_Gx^lx^m,
\label{fnmetric}
\end{equation}
where Latin indices are taken to be spatial, and it is to be noted that the curvature components are evaluated on $G$, where $x^i=0$, i.e on the geodesic, where the metric is that of
flat space-time (with Lorentzian signature), $\eta_{\alpha\beta}= \left(-1,1,1,1 \right)$. Also note that the dependence on the observer's time enters the metric only through the 
components of the curvature that are evaluated at a given value of the proper time along the geodesic. Now
having set up such a coordinate system, we can 
analyze the covariant Dirac Lagrangian 
\begin{equation}
{\mathcal L} = \sqrt{-g}\left(i{\bar \psi}\gamma^{\alpha} D_{\alpha} \psi - m{\bar \psi}\psi\right),
\label{Dirac}
\end{equation}
where $\gamma^{\alpha}$ are the usual Dirac matrices. First, a word about the notation is in order. We have three types of space-time metrics at play here: the original space-time that we start from, a 
locally flat metric at the fermion and finally a curved space-time around the geodesic given by the metric of Eq.(\ref{fnmetric}). Following our previous notation, 
the primed coordinates will always refer to the original space-time. We will use the beginning Greek indices $\alpha, \beta, \cdots$ to denote the locally flat metric. 
The later Greek indices like $\mu, \nu \cdots$ will be used to denote the curved space-time surrounding our geodesic. 

In Eq.(\ref{Dirac}), the covariant derivative and the spin connection have the standard definitions 
\begin{equation}
D_{\alpha} = \left(\partial_{\alpha} - \frac{i}{4} \omega_{\beta\gamma\alpha}\sigma^{\beta\gamma}\right), ~~~\omega_{\alpha\beta\gamma} = 
e_{\alpha\mu}\left(\partial_\gamma e^{\mu}_\beta + \Gamma^{\mu}_{\nu\rho}e^{\nu}_\beta e^{\rho}_\gamma\right),~~~\sigma^{\alpha\beta}=
\frac{i}{2}[\gamma^\alpha,\gamma^\beta],
\end{equation}
where $\Gamma^{\mu}_{\nu\rho}$ is a Christoffel connection, and $e^{\mu}_\alpha$ denotes a tetrad basis connecting the curved indices and flat indices near the geodesic.
It can be shown \cite{mohanty}, \cite{bani} that the terms which come
from the spin connection involve an interaction Lagrangian of the form ${\bar \psi}\gamma^{\alpha}\gamma^5b_{\alpha}$, where the four vector $b$ can be written in a compact form :
\begin{equation}
b^\sigma = \epsilon^{\alpha\beta\gamma\sigma}e_{\beta\mu}\left(\partial_\alpha e^{\mu}_k + \Gamma^{\mu}_{\nu\rho}e^{\nu}_\gamma e^{\rho}_{\alpha}\right) \equiv  
\epsilon^{\alpha\beta\gamma\sigma}e_{\beta\mu}\partial_\alpha e^{\mu}_\gamma,
\label{bs}
\end{equation}
where $\gamma^5 = i\gamma^0\gamma^1\gamma^2\gamma^3$, and it can be checked that $b_0$ is identically zero.
Importantly, in a Hamiltonian approach, this interaction Lagrangian can be cast into an effective interaction energy of the form $-{\vec b}.{\vec s}$ in the non-relativistic limit, as shown in 
\cite{kostelecky2} (see also \cite{kostelecky1}). This involves starting from a relativistic Dirac Lagrangian, and extracting a non-relativistic quantum Hamiltonian by following 
a sequence of Foldy-Wouthuysen transformations with appropriate field redefinitions. The fact that the interaction Lagrangian of Eq.(\ref{bs}) can be cast into this non-relativistic
form is the basis of gravitation induced CPT violation in the EOT-Wash experiment. In our case, as explained in \cite{mohanty}, the magnetic fields ${\vec b}$ will change sign under
parity, and hence gravitational curvature couplings will not violate CPT. These magnetic fields might however form the basis of other detectable gravitational effects. 

In order to compute the magnetic field, we will need the tetrads $e^{\mu}_{\nu}$, and these can be shown to be \cite{parker1}, 
\cite{parker3}
\begin{equation}
e^{\mu}_0 = \delta^{\mu}_0 - \frac{1}{2} R^{\mu}_{~\alpha 0 \beta}|_G~x^{\alpha}x^{\beta},~~~e^{\mu}_i = \delta^{\mu}_i - \frac{1}{6}R^{\mu}_{~jik}|_G~x^jx^k,
\label{tetradsfn}
\end{equation}
where $i$, $j$ and $k$ run over the spatial indices only. Hence, to calculate the $b_i$'s which is one of our main interests in this paper, we require to set up the tetrad basis 
of Eq.(\ref{tetradmain}) to evaluate the components of the Riemann tensor in Fermi normal coordinates. We then use this input in Eq.(\ref{bs}) along with the tetrads in
Eq.(\ref{tetradsfn}) to obtain the fields. The
effect of gravity will enter the field through the curvature tensor. As we have mentioned, $b_0$ vanishes identically, and the expression for the spatial 
components ${\vec b}$ are given by \cite{mohanty}
\begin{equation}
b_i = \frac{1}{4}\epsilon_{0\gamma\beta i}R^{\gamma\beta}_{~~0l}|_G~x^l + \frac{1}{4}\epsilon_{0\gamma\beta i}R^{0\beta\gamma}_{~~~l}|_G~x^l,
\label{magfield}
\end{equation}
Hence, knowing $R_{\alpha\beta\gamma\delta}$, we can compute the effective magnetic field in the non-relativistic limit due to gravitational interactions, and in cases of interest, this should be contrasted
with the intrinsic magnetic fields present in some celestial objects. 

One has to be careful with the coordinates of the event here, since the metric in Fermi normal
coordinates of Eq.(\ref{fnmetric}) is valid only close to the geodesic. If the measurement is carried out on the geodesic where $x^l = 0$, ${\vec b}$ vanishes 
identically. Typically, in computations related to the Hydrogen atom (which is in free fall) for example, the hypothetical observer is located at the nucleus of the atom, and the $x^i$s 
are hence of the order of the Bohr radius \cite{parker1}. For the purpose of our discussion, we will keep
the observer's coordinate explicit, and in some cases where we discuss quantitative results, we will take this to be of order unity, since we are mainly interested in experiments
involving finite size apparatus. 

A few words about the magnitudes of ${\vec b}$ is also in order. Present day bounds on the effective magnetic fields can be obtained from the fact that ${\vec b}$ translates to a magnetic field
${\vec B} = {\vec b}/\mu_B$ where $\mu$ is the Bohr magneton. Using $\mu_B = 9.3 \times 10^{-24} {\rm J/T}$, the fact that ${\vec B}$ can be measured upto an accuracy of 
$10^{-12} {\rm Gauss}$ translates into $|{\vec b}| \sim 10^{-28} - 10^{-29} {\rm GeV}$  \cite{kostelecky1}, \cite{kostelecky3}. Hence, if we in a realistic scenario, ${\vec b}$ turns out to be
of this order, then we can hope to detect it. However, as we point out in sequel, the values of the magnetic fields that we obtain are much less than present day bounds, and hence 
we can only hope that these will be detected in futuristic experiments. However, we do not rule out the fact that indirect evidence of strong magnetic fields due to gravitational effects
might be possible to observe. 

Before we begin our analysis, let us point out what we expect. On physical grounds, we expect that large ${\vec b}$ should occur in regions where gravitational 
effects are large and our observer is close to an instability. For circular geodesics, vacuum solutions of Einstein's equations dictate that such instabilities can occur near a photon
sphere \cite{ve} which is defined as a time-like hypersurface, such that a null geodesic that is tangent to this hypersurface at some point of time will remain so in future. 
For the Schwarzschild black hole, the photon sphere is at $r=\frac{3}{2}R_s$, with $R_s$ being the Schwarzschild radius, and at this value of the radial coordinate, the energy and
angular momentum per unit mass of a test particle undergoing circular geodesic motion becomes very large, i.e the motion is highly relativistic. Indeed, it is well known that scalar-field 
power spectrum from particles in circular geodesics in a Schwarzschild background sharply peaks near the photon sphere \cite{misnerpower}. We expect that ${\vec b}$ should be
large in this limit for Schwarzschild and for similar limits in the JNW space-times as well, where behavior of circular geodesics is qualitatively similar to (but has a richer structure than) Schwarzschild backgrounds.
Here, we run into an important issue of stability of orbits. Circular orbits near the photon sphere in exterior Schwarzschild backgrounds are unstable, and hence the physics of magnetic
fields for such orbits is somewhat unclear. However, as we will show, there are stable orbits for which the fields will be large, and in these cases the physicality of our results is guaranteed.
Also, for BSTs that we consider later, there is no photon sphere. Here, stable circular orbits are possible close to the central singularity, by construction. In this case, we expect that the magnetic 
field becomes large only close to the singularity. We will indeed see that these expectations are met. 

For radial geodesics, the situation is different. Here, in general we would expect the field ${\vec b}$ to depend on the energy of the observer, since generically this would appear in
the curvature (an important exception being the Schwarzschild solution). Hence, in this case, the results will depend on the observer's velocity. More energetic observers should
feel the effect of gravity more than less energetic ones, and this is what we also find from our results. 

\subsection{Fermi Normal Coordinates for Circular Geodesics in Static, Spherically Symmetric Space-times} 

We consider a general static, spherically symmetric space-time with metric\footnote{We will keep the factors of the speed of light $c$
explicit for the moment, so that dimensional consistency of our results can be verified at each stage. For qualitative discussions, we will set $c=1$.}  :
\begin{equation}
 ds^2 = -c^2A(r)dt^2 + B(r)dr^2 + G(r)d\Omega^2
\label{genmetric}
\end{equation}
where $A(r)$, $B(r)$ and $G(r)$ may be arbitrary positive functions of the radial coordinate, and $d\Omega^2 = d\theta^2 + \sin^2\theta d\phi^2$ is the standard metric on the
unit two sphere. The form of the metric dictates that we have the conserved quantities
\begin{equation}
\epsilon = c^2A(r){\dot t},~~~L = G(r){\dot \phi},
\end{equation}
which are the energy per unit mass and angular momentum per unit mass of the test particle, respectively. Here and in what follows, the dots denote derivatives with respect 
to the proper time on the geodesic and the primes will denote a derivative with respect to $r$. For time-like geodesics, we have
\begin{equation}
{\dot r}^2 + V(r) = 0,~~~V(r) = \frac{1}{B(r)}\left[-\frac{\epsilon^2}{c^2A(r)} + \frac{L^2}{G(r)} + c^2\right].
\end{equation}
For circular orbits, using $V(r) = V'(r) = 0$, we obtain 
\begin{equation}
\epsilon = \frac{c^2 A(r)\sqrt{G'(r)}}{\sqrt{G'(r)A(r)-G(r) A'(r)}} ,~~~L = \frac{c G(r) \sqrt{A'(r)}}{\sqrt{A(r) G'(r)-G(r) A'(r)}}
\end{equation}
Now one can set up the Fermi normal tetrad basis as follows (the Schwarzschild case was worked out in \cite{parker3})
\begin{eqnarray}
{\hat e}^{\mu'}_{0} &=&\left(\frac{\epsilon }{c^3 A(r)}, 0, 0, \frac{L}{c G(r)}\right),~~~\nonumber\\
{\hat e}^{\mu'}_{1}&=&\left(-\frac{L \sin (\phi (\tau )\delta(r))}{c^2 \sqrt{A(r) G(r)}}, 0, \frac{\cos
(\phi (\tau )\delta(r))}{\sqrt{B(r)}} , 0 , -\frac{\epsilon  \sin (\phi (\tau)\delta(r))}{c^2 \sqrt{A(r) G(r)}}\right),\nonumber\\
{\hat e}^{\mu'}_{2} &=&\left( 0 , 0 , \frac{1}{\sqrt{G(r)}} , 0\right),~~~\nonumber\\
{\hat e}^{\mu'}_{3} &=&\left(\frac{L \cos (\phi (\tau )\delta(r))}{c^2 \sqrt{A(r) G(r)}} , \frac{\sin (\phi (\tau
)\delta(r))}{\sqrt{B(r)}} , 0 , \frac{\epsilon  \cos (\phi (\tau )\delta(r))}{c^2 \sqrt{A(r) G(r)}}\right),
\label{gentetrad}
\end{eqnarray}
where we have defined 
\begin{equation}
\delta(r)=\frac{1}{2} \sqrt{\frac{G'(r) \left(A(r) G'(r)-G(r)A'(r)\right)}{A(r) B(r) G(r)}}.
\end{equation}
For example, for the Schwarzschild black hole, this factor is $\delta(r) = \sqrt{1-\frac{3R_s}{2 r}}$, where $R_s =\frac{2GM}{c^2}$ is the Schwarzschild radius and $G$ denotes the Newton's constant. 

It can be checked that the form of the tetrad above satisfies the conditions in Eq.(\ref{tetradmain}). Now we can compute the components of the curvature tensor in the coordinates
of Eq.(\ref{genmetric}) and use Eqs.(\ref{Rfr}) and (\ref{gentetrad}) to obtain the curvature compoents in Fermi normal coordinates. The calculations are lengthy, but can
be easily performed using a standard Mathematica routine. The expressions are large and we omit them 
for brevity. We use Eq.(\ref{magfield}) to evaluate the effective magnetic field due to curvature couplings, and find that 
\begin{eqnarray}
 b_{0} &=& 0,~~~b_1 = {\sqrt{A'G'}[GA'G'+A(4BG-G'^2)]\over
16BA^{1/2}G^{3/2}(AG'-GA')}y\cos(\phi(\tau) \delta)\nonumber\\
b_2 &=& \frac{P(r)}{Q(r)},~~~b_{3}={\sqrt{A'G'}[GA'G'+A(4BG-G'^2)]\over
16BA^{1/2}G^{3/2}(AG'-GA')}y\sin(\phi(\tau) \delta)
\label{general}
\end{eqnarray}
where $P(r)$ and $Q(r)$ are defined as
\begin{eqnarray}
 P(r) &=& \sqrt{A'G'}\Big[B\Big(2A^2GG''-A^2G'^2+G^2(A'^2-2AA'')\Big)\nonumber\\
 &+& AGB'(GA'-AG')\Big]\Big(x\cos (\phi(\tau)\delta)+z \sin (\phi(\tau)\delta)\Big)
\nonumber\\
 Q(r) &=& 16B^2(AG)^{3/2}(AG'-GA'),
\end{eqnarray}
and $x,y,z$ are the observer's spatial coordinates. If $x^i=0$, i.e the measurements are carried out on the geodesic, the magnetic field is zero as expected,
as Fermi normal coordinates are flat there. In what follows, we will present our results by setting $x^i \sim O(1)$. In graphical analysis, we will calculate the fields
in units of the observer's typical length, which we denote by $R_o$. In spherical polar coordinates, this will be the radius of the event. 

Eq.(\ref{general}) is the master equation for this section, and can be used to compute the effective magnetic field for any static, spherically symmetric metric. 
Having outlined the general calculation scheme, we will now specialize to our cases of interest. The first case we consider is fermions in circular motion in the 
background of the Schwarzschild solution. 

\subsection{Circular Geodesics in Exterior and Interior Schwarzschild Space-times}

\begin{figure}[t!]
\begin{minipage}[b]{0.5\linewidth}
\centering
\includegraphics[width=2.8in,height=2.3in]{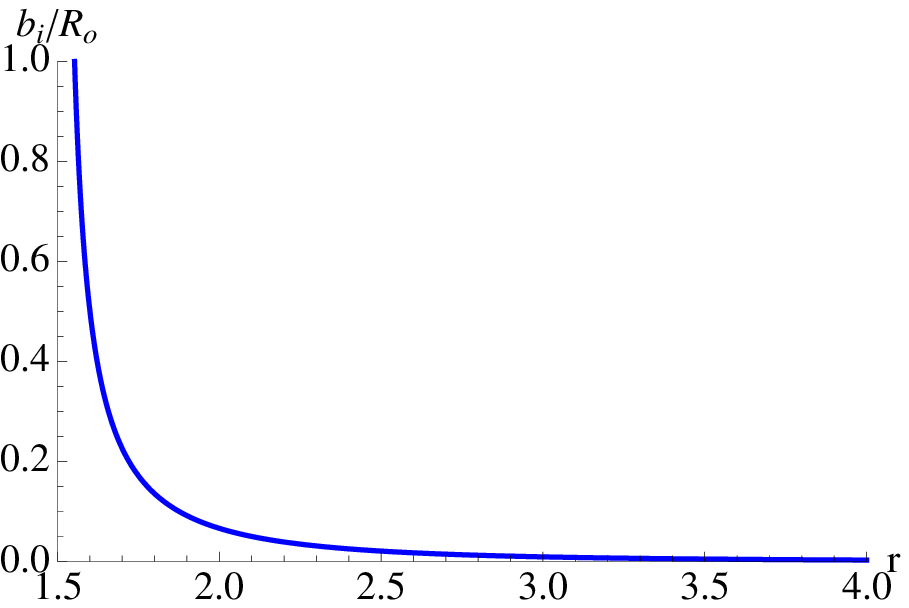}
\caption{Color Online : Effective magnetic field in units of the observer's length scale, as a function of distance for the exterior Schwarzschild solution, with $R_s=1$.}
\label{SchE}
\end{minipage}
\hspace{0.2cm}
\begin{minipage}[b]{0.5\linewidth}
\centering
\includegraphics[width=2.8in,height=2.3in]{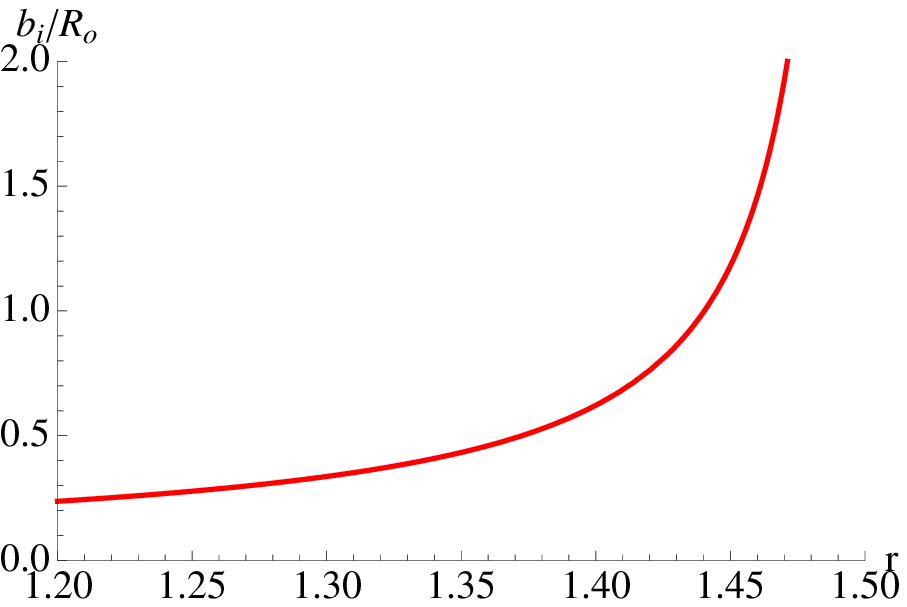}
\caption{Color Online : Effective magnetic field in units of the observer's length scale, for the interior Schwarzschild solution, with $R_s=1$, $R=1.5$. }
\label{SchI}
\end{minipage}
\end{figure}

We start with the familiar exterior Schwarzschild metric 
\begin{equation}
ds^2 = -\left(1 - \frac{R_s}{r}\right)c^2dt^2 + \left(1 - \frac{R_s}{r}\right)^{-1}dr^2 + r^2d\Omega^2,
\end{equation}
The energy and angular momentum
per unit mass of a particle in circular geodesic is given in the Schwarzschild background as 
\begin{equation}
\epsilon = \frac{c^2 \left(1-\frac{R_s}{r}\right)}{\sqrt{1-\frac{3 R_s}{2r}}},~~~L = \sqrt{\frac{c^2 r^2 R_s}{2 r-3R_s}}
\end{equation}
In this case, our calculation yields, from the master Eq.(\ref{general})\footnote{We list in Appendix A the non-zero components of the Riemann tensor for circular geodesics in
Schwarzschild backgrounds, also obtained in \cite{collas}.}
\begin{eqnarray}
b_1 &=& {\mathcal A} y \cos \left(\phi (\tau ) \delta\right)\nonumber\\
b_2 &=& {\mathcal A}\left(x\cos \left(\phi (\tau ) \delta\right)+z\sin \left(\phi (\tau ) \delta\right)\right)\nonumber\\
b_3 &=& {\mathcal A}y\sin \left(\phi (\tau ) \delta\right),
\label{bSch}
\end{eqnarray}
where we have defined 
\begin{equation}
{\mathcal A} = \frac{3\sqrt{R_s^3 (r-R_s)}}{4 \sqrt{2} r^3 (2 r-3R_s)}.
\label{schcmain}
\end{equation}
Let us analyze this result in some details.\footnote{For ease of notation, we will henceforth write the magnetic fields in units of the observer's coordinates, and ignoring
the angular factor. This will not lead to any loss of generality of our results, while avoiding cluttering notations. Thus, we will calculate $b_i/R_o$, where $R_o$ is a typical
coordinate of the observer defined at the end of section 2.2. However, we will keep denoting this as $b_i$. The full expression for the fields will typically look like that in Eq.(\ref{bSch}).}
First, we note from Eq.(\ref{bSch}) that the magnetic field blows up close to $r = \frac{3}{2}R_s$, the location of the photon sphere 
(below this radius, circular orbits do not exist). As alluded to earlier in this section, although the circular orbit is unstable at this radius, 
this implies that gravitational contribution to the magnetic field can be very large here. To get a physically meaningful situation, we should have
an object whose radius is of the order of its Schwarzschild radius. A neutron star is a candidate which fits this requirement, and hence our result is indicative of the
fact that near the photon sphere of a neutron star, GR effects can generate a very large magnetic field on a fermion in circular motion. Of course, as said in the introduction, this is to be distinguished
from the intrinsic magnetic fields of such objects whose physics is very different. The field falls off as $\sim r^{-\frac{7}{2}}$ as a function of the radial 
distance, for large $r$. In fig.(\ref{SchE}), we plot the magnetic field as a function of the 
distance, with $R_s = 1$. As a numerical estimate, upon restoring factors of $\hbar$ and setting $R_s = 10^3 m$ (a typical Schwarzschild radius for a neutron star) we obtain
$|{\vec b}| \sim  10^{-24}$ GeV at a distance $r=3R_s$, where stable orbits exist. 
We also note that in the direction of the magnetic field, given say in the $x-z$ plane of the observer by ${\rm tan}^{-1}\left(\frac{b_3}{b_1}\right)$, the $r$ dependence
enters only via $\delta(r)$. Far from the black hole, this quantity is small, and hence in this regime, the direction of the magnetic field will be independent of the Schwarzschild radial coordinate $r$.  

In is instructive to contrast the results above with those obtained from an interior Schwarzschild solution. To simplify notations, we will define
\begin{equation}
\alpha = \frac{3}{2} \sqrt{1-\frac{R_s}{R}},~~~\kappa = \frac{R_s}{R^3},~~~X = \left(\alpha - \frac{1}{2}\sqrt{1-\kappa r^2}\right)^2.
 \label{alphabeta}
 \end{equation}
 Then the well known interior Schwarzschild solution, describing a fluid of constant density, is given by 
 \begin{equation}
 ds^2 = -c^2Xdt^2 + \frac{dr^2}{1-\kappa r^2} + r^2d\Omega^2,
 \label{extSch}
 \end{equation}
the solution being valid for $r \leq R$, $R$ being the matching radius where the interior solution goes over to an external Schwarzschild one. 
Here, we will use the conserved energy and angular momentum per unit mass, given as
\begin{equation}
\epsilon = c^2\sqrt{\frac{ \sqrt{1-\kappa  r^2} \left(\sqrt{1-\kappa  r^2}-2 \alpha
   \right)^3}{4 \left(1 - 2\alpha\sqrt{1-\kappa r^2}\right)}},~~~L = cr^2\sqrt{\frac{\kappa}{2 \alpha \sqrt{1-\kappa r^2}-1}}
   \end{equation}
respectively. In this case, general results for the magnetic field derived from our master Eq.(\ref{general}) is lengthy and not very illuminating. We find that for small values of 
$r$, the fields start from small values, and their magnitudes increase as $r$ increases. This is expected, as
our observer experiences more gravitating mass as the radial distance increases, and hence the effective magnetic field also increases. 
Thus, the magnetic field will be strongest at the matching radius. At this value $r=R$, the expressions for the fields simplify, and we get 
\begin{equation}
b_1 = b_3 = -b_2 = \frac{3\sqrt{R_s^3(R - R_s)}}{4\sqrt{2}R^3\left(2R - 3R_s\right)}.
\end{equation}
Note the similarity of this expression with the results obtained from the exterior Schwarzschild solution (Eqs.(\ref{bSch}), (\ref{schcmain})). The solutions match but for 
a negative sign in $b_2$. Of course, this is insignificant, since the fields depend on the derivatives of the metric, and these need not be continuous at the matching radius. 
We see however that in this case also, a large magnetic field might be generated in the interior region, if ($r=$) $R = \frac{3}{2}R_s$, but importantly, circular orbits are stable here,
i.e, at $r=R$, the conserved energy and angular momentum per unit mass blows up, i.e the
circular orbit is highly relativistic, but is stable as can be checked by calculating the effective potential. 
Typically, such a situation might again be of relevance
in the context of highly gravitating objects like black holes or neutron stars. In fig.(\ref{SchI}), we show this result graphically. Here we have set $R_s = 1$ and $R = 1.5$.
 
\subsection{Circular Geodesics in JNW space-times}

Next, we move on to consider circular geodesics in the Janis-Newman-Winicour space-times. These are space-times sourced by a scalar field, i.e this is an
Einstein-Klein-Gordon systems, given by the metric 
 \begin{equation}
ds^2_{\rm JNW} = -c^2\left(1-\frac{B}{r}\right)^\nu dt^2 + \frac{1}{\left(1-\frac{B}{r}\right)^\nu}dr^2 + r^2\left(1-\frac{B}{r}\right)^{1-\nu}d\Omega^2.
\label{a1}
\end{equation}
Here, $\nu$ is a parameter that ranges from $0$ to $1$. As is well known, the singularity of this space-time at $r=B$ is globally naked. 
Also, the source of the JNW space-time is a scalar field,
\begin{equation}
 \psi = \frac{q}{B\sqrt{4\pi}}\ln\left(1-\frac{B}{r}\right)
 \label{a2}
\end{equation}
where $q$ is a parameter that denotes its magnitude. The ADM mass $M$ is related to the $B$ and $q$ by 
$B = 2\sqrt{q^2 + M^2}$, and also $\nu = 2M/B$.\footnote{We will set $G=c=1$ here. Setting $B=1$ fixes $M=\frac{\nu}{2}$ for a given value of $\nu$. This sets the value of $q$
via $q^2 = \frac{1}{4} - M^2$. }
Setting $\nu = 1$, i.e $q=0$, one recovers back the Schwarzschild metric. For JNW space-times, 
we use the conserved energy and angular momentum per unit mass as 
\begin{equation}
\epsilon = c^2\sqrt{\left(1-\frac{B}{r}\right)^{\nu } \left(1+\frac{B \nu }{2r-2 B \nu -B}\right)},~~~L = cr\sqrt{\frac{B\nu\left(1-\frac{B}{r}\right)^{1-\nu }}{2r-2 B \nu -B}}.
\label{jnwea}
\end{equation}
\begin{figure}[t!]
\begin{minipage}[b]{0.5\linewidth}
\centering
\includegraphics[width=2.8in,height=2.3in]{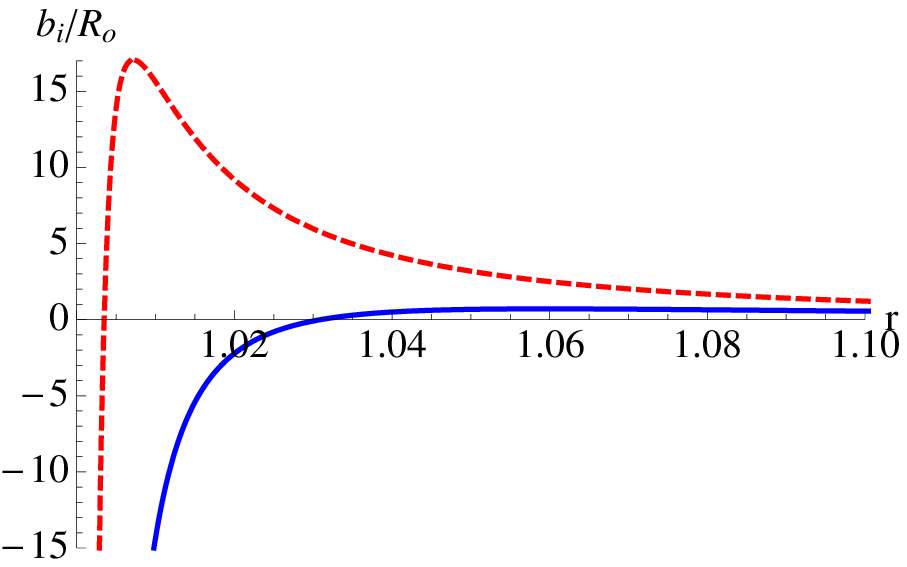}
\caption{Color Online : Effective magnetic fields as a function of the radial distance for JNW space-time with $\nu = 0.43$ (solid blue) and $\nu = 0.49$ (dashed red).}
\label{jnw1}
\end{minipage}
\hspace{0.2cm}
\begin{minipage}[b]{0.5\linewidth}
\centering
\includegraphics[width=2.8in,height=2.3in]{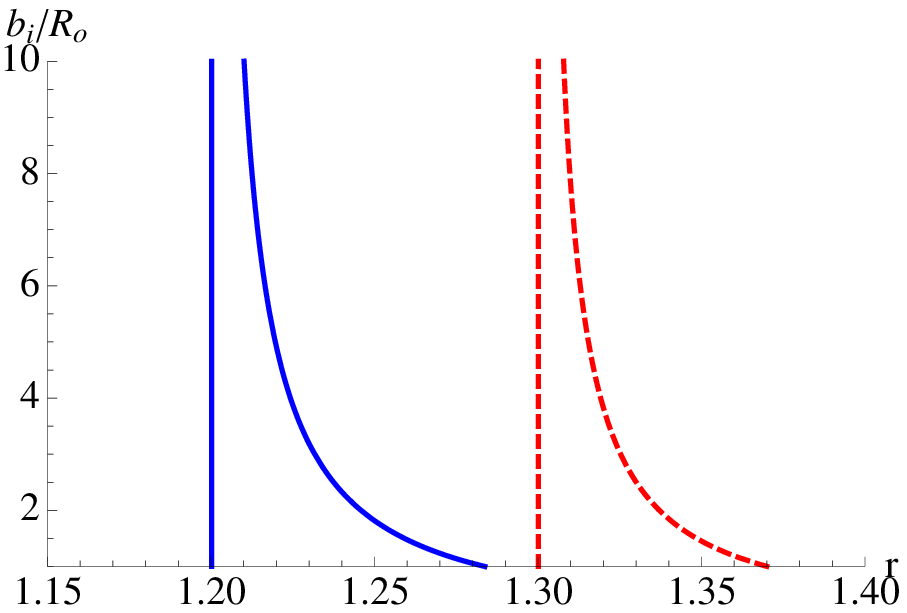}
\caption{Color Online : Effective magnetic fields as a function of the radial distance for JNW space-time with $\nu = 0.7$ (solid blue) and $\nu = 0.8$ (dashed red).}
\label{jnw2}
\end{minipage}
\end{figure}

Now we can calculate the magnetic fields as before. For illustration, we will set $B=1$, in which case the fields take a simple form 
\begin{equation}
b_1=b_2=b_3 = \frac{(r-1)^{\nu-2}r^{-\nu-2}\left(\nu(6 r-3)-2\nu^2-1\right) \sqrt{\nu (2r - \nu - 1)}}{16 (2r-2\nu-1)}
\label{jnwc}
\end{equation}
For $\nu \to 1$, this goes over to the Schwarzschild result, as expected. To analyze Eq.(\ref{jnwc}), we first note that the photon sphere for the JNW singularity is at 
\begin{equation}
r_{\rm ps} = \frac{B}{2}\left(1 + 2\nu\right).
\end{equation}
This is the radius at which the energy and angular momentum per unit mass of Eq.(\ref{jnwea}) diverges. It is also known from an analysis of the stability of circular
orbits that three distinct regions in $\nu$ need to be considered. Defining 
\begin{equation}
r_{\pm} = \frac{B}{2}\left(1+3\nu \pm \sqrt{5\nu^2 - 1}\right),
\end{equation}
the ranges of stable orbits corresponding to different values of $\nu$ are (see, e.g. \cite{drs}) 
\begin{eqnarray}
&~& {\rm Case~1 :}~~~0 < \nu < \frac{1}{\sqrt{5}},~~~ B < r < \infty;\nonumber\\
&~&{\rm Case~2 :}~~~\frac{1}{\sqrt{5}} < \nu < \frac{1}{2},~~~B < r < r_- \hspace{4mm} \mbox{and} \hspace{4mm}  r_+ < r < \infty; \nonumber\\
&~&{\rm Case~3 :}~~~\frac{1}{2} < \nu < 1,~~~ r_+ < r < \infty.
\end{eqnarray}
Now from Eq.(\ref{jnwc}), we see that the magnetic field becomes very large near the photon sphere, whenever it exists. Specifically, from this equation,
$b_i$ diverges at $r = r_{\rm ps} = \frac{1}{2}(1+2\nu)$ and at $r=B=1$. Let us first consider case I. As a representative value, we choose $\nu=0.43$. 
Here, stable orbits exist at all radii. In fig.(\ref{jnw1}), we have shown
this case with the solid blue line, where the field is plotted as a function of the radial distance $r$. They diverges close to the singularity at $r=1$. Next, we consider
case 2, where we have chosen $\nu=0.49$. Stable orbits exist for $r < 1.01$ and for $r> 1.46$. The dashed red curve of fig.(\ref{jnw1}) depicts the fields in this
case. We see that there is a divergence in the fields at $r=1$. There is a maximum of the field at $r=1.007$, where in units of the observer's coordinates, $b_i \sim 17$ 
(in appropriate energy units). At both these values of $r$,
the orbits are stable. For case 3, we find that the fields become large only near the photon sphere, where circular orbits are highly relativistic, but unstable. In 
fig.(\ref{jnw2}), we show this situation for $\nu = 0.7$ (solid blue curve) and for $\nu = 0.8$ (dashed red curve). The radii of the photon sphere are at 
$r=1.2$ and $r=1.3$ respectively. 

\subsection{Circular Geodesics in Bertrand Space-times}

We will now consider circular geodesics in BSTs. First let us recall a few definitions. BSTs were discovered by 
Perlick \cite{perlick}, as solutions of Einstein gravity where each spatial point admits a closed stable orbit. This generalizes the well known Bertrand's theorem \cite{goldstein}
to GR. That such a space-time can be good candidates for galactic dark 
matter was pointed out in \cite{dbs1} from the assumption that stars away from galactic centres follow approximately circular orbits. The metric for a BST (of type II, in the
classification of \cite{perlick}) is given by
\begin{eqnarray}
ds_{\rm BST}^2 = -\frac{c^2dt^2}{D +\frac{\alpha}{r}} + \frac{dr^2}{\beta^2} + r^2 d\Omega^2\,.  
\label{type2a}
\end{eqnarray}
Here, $\alpha$ and $D$ are positive, $\beta$ is a rational number, and $r_s = \frac{\alpha}{D}$ is related to the galactic length scale, in the sense that if we take $r_s$ to be of the size of the galaxy,
then in the Newtonian limit, reasonable estimates to the mass of the galaxy can be obtained. Indeed, a phenomenological definition of the circular velocity 
of matter (considered as a perturbation over the BST background such that back reaction effects are neglected) gives results that match well with experimental data
on low surface brightness galaxies. In a Newtonian approximation, these also reproduce the popular Navarro-Frenk-White (NFW) \cite{nfw} and Hernquist \cite{hernquist} density
profiles for the dark matter distribution, in appropriate limits \cite{dbs1}. We will consider this model in a phenomenological spirit and estimate effective magnetic fields
for observers in geodesic motion in this matter distribution, even close to the galactic centre. Note that there is a central singularity at $r=0$ in the metric of Eq.(\ref{type2a})
which is naked. 

Here, we evaluate ${\vec b}$ by setting $\beta = 4/5$. \footnote{Non-zero components of the Riemann tensor for circular geodesics in BSTs are listed in Appendix B.}
\begin{eqnarray}
b_1 &=& \frac{\alpha  (17 \alpha +9 D r)}{50 \sqrt{2} r^2 \sqrt{\alpha  (\alpha +D r)} (\alpha +2 D r)} = b_3\nonumber\\
b_2&=& \frac{\sqrt{2} \left(\frac{\alpha }{\alpha +D r}\right)^{3/2} (\alpha +4 D r)}{25 r^2 (\alpha +2 D r)}
\label{magbst}
\end{eqnarray}   
First of all, we note from the above expressions that for small distances $Dr \ll \alpha$, $b_i \sim r^{-2}$. For large distances, we have $b_1$ and $b_3$
$\sim r^{-\frac{5}{2}}$ while $b_2 \sim r^{-\frac{7}{2}}$. Specifically, in the first case, we see that $b_i$ becomes independent of $\alpha$ and $D$, and $\sim r^{-2}$,
whereas in the opposite limit, we obtain $b_1, b_3 = K_1 r_s^{\frac{1}{2}}r^{-\frac{5}{2}}$ and $b_2 = K_2r_s^{\frac{3}{2}}r^{-\frac{7}{2}}$, where $K_1$ and $K_2$ are
numerical constants. The fact that the fields blow up as $r\to 0$ is expected, since the energy density in BSTs also blow up in this limit \cite{dbs1}. 
Also, we see that the direction of the magnetic field in one of the planes of the observer (the $X-Y$ plane in this case) is dependent on $r$,
\begin{equation}
\frac{b_2}{b_1} = \frac{4 \alpha  (\alpha +4 D r)}{(\alpha +D r) (17 \alpha +9 D r)}
\end{equation}
a result that is qualitative different from a Schwarzschild background. 

Now, we note that from a Newtonian perspective, a fit of the circular velocities of galactic rotation curves relates the parameters $\alpha$ and $D$ to the
maximum value of the circular velocity and the Newtonian mass of the galaxy as (Eqs.(8) and (12) of \cite{dbs1})
\begin{equation}
\alpha = \frac{c^2 G M}{8 (v_c^{\rm max})^4},~~~D=\frac{c^2}{8 (v_c^{\rm max})^2}
\label{bstalphad}
\end{equation}
where $\frac{\alpha}{D}$ can be taken to be an estimate of the size of the galaxy. If we input these values in the expressions for the magnetic field of Eq.(\ref{magbst}), we obtain
\begin{eqnarray}
b_1 &=& b_3 = \frac{17 G M+9 r (v_c^{\rm max})^2}{50 r^2 \left(G M+2 r (v_c^{\rm max})^2\right) \sqrt{2+\frac{2r (v_c^{\rm max})^2}{G M}}}\nonumber\\
b_2 &=& \frac{\sqrt{2} \left(\frac{G M}{G M+r (v_c^{\rm max})^2}\right)^{3/2} \left(G M+4 r(v_c^{\rm max})^2\right)}{25 r^2 \left(G M+2 r (v_c^{\rm max})^2\right)} 
\end{eqnarray}
Now we can take a typical estimate of $M \sim 10^8M_{\odot}$ and $v_c^{\rm max} = 30 {\rm km/sec}$ (which describe the galaxy NGC4395 to a good approximation), 
to obtain $b_i \sim 10^{-26} {\rm GeV}$ for $r \sim 10^4 {\rm m}$. \footnote{Upon restoring all units, the $b_i$'s are of dimension $L^{-1}$. This has to be multiplied by $\hbar c$
in order to get the field in GeV.} This is of course a somewhat unrealistic estimate, since we have taken $r$ to be very small compared to the 
galactic scales, but might be observationally important. If we take $r = \alpha/D$, which is an estimate for the galactic size, we obtain 
\begin{equation}
b_i \sim \frac{(v_c^{\rm max})^4}{G^2M^2},
\end{equation}
and taking the typical values of $M$ and $v_c^{\rm max}$ indicated above, we obtain $b_i \sim 10^{-55}{\rm GeV}$ which is too small to detect in present experiments. 

\section{Radial Geodesics and Fermion Curvature Couplings}

We now turn to observers in radial geodesics as described by Fermi normal coordinates. We will first demonstrate the general construction of Fermi normal coordinates for 
radial motion in static, spherically symmetric space-times, and present the results for the effective magnetic field in the general case. Then we will specialize to some examples. 

\subsection{Fermi Normal Coordonates for Radial Geodesics in Static, Spherically Symmetric Space-times}

We start with the general metric of Eq.(\ref{genmetric}). For radial geodesics, we set up the tetrad
\begin{eqnarray}
{\hat e}^{\mu'}_{0} &=& \left(\frac{{\dot t}}{c}, \frac{{\dot r}}{c}, 0, 0\right),~~~{\hat e}^{\mu'}_{1} = \left(\frac{{\dot r}}{c^2}\sqrt{\frac{B(r)}{A(r)}}, {\dot t}
\sqrt{\frac{A(r)}{B(r)}},0,0\right)\nonumber\\
{\hat e}^{\mu'}_{2} &=& \left(0,0,\frac{1}{\sqrt{G(r)}},0\right),~~~{\hat e}^{\mu'}_{3} = \left(0,0,0,\frac{1}{\sqrt{G(r)}{\rm sin}\theta}\right),
\label{radialfn}
\end{eqnarray}
where the dot denotes a derivative with respect to the proper time along the radial geodesic. Now one can check that the conditions of Eq.(\ref{tetradmain}) are satisfied, upon using the 
radial geodesic equation in this background. We also use the normalization condition for radial geodesics 
\begin{equation}
{\dot t} = \sqrt{\frac{c^2 + B(r){\dot r}^2}{c^2A(r)}} = \frac{\epsilon}{c^2A(r)},
\end{equation}
where $\epsilon$ is the conserved energy per unit mass of our test particle, as before. Now we can compute the effective magnetic field. We find that
$b_0=b_1=0$, and 
\begin{equation}
b_2 = z{\mathcal B}, ~~ -b_3 =y{\mathcal B},~~{\mathcal B}= 
\frac{1}{16c B G^2}z \dot r \sqrt{\frac{B(r)\dot
r^2+c^2}{A(r)c^2}}\Big(M(r)+N(r)\Big).
\label{mainradial}
\end{equation}
where we have defined
\begin{eqnarray}
M(r) &=& \sqrt{A \over B}B G'^2 \nonumber \\
N(r) &=& G\Big(\sqrt{\frac{B}{A}}A' G'+\sqrt{\frac{A}{B}}(B'G'-2BG'')\Big)
\end{eqnarray}
Eq.(\ref{mainradial}) is the master equation for this section, and we will now proceed to analyze special cases of this result. 

\subsection{Radial Geodesics in Schwarzschild, JNW and Bertrand Space-times}

\begin{figure}[t!]
\centering
\includegraphics[width=2.8in,height=2.3in]{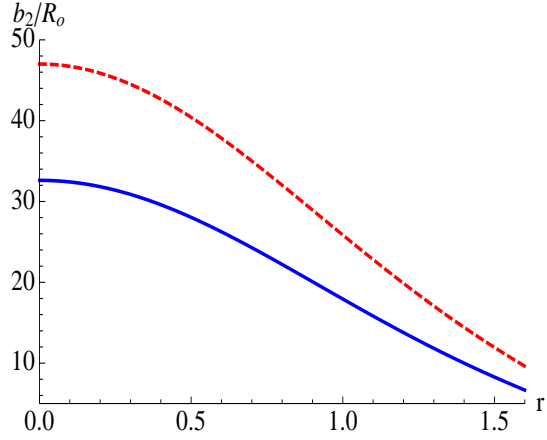}
\caption{Color Online : Effective magnetic fields as a function of the radial distance for the interior Schwarzschild space-time with $\epsilon = 5c^2$ (solid blue) and $\epsilon = 6c^2$ (dashed red).}
\label{SchIr}
\end{figure}

For the exterior vacuum Schwarzschild solution, we set up the Fermi normal coordinates as prescribed in \cite{ferminormal} (or from Eq.(\ref{radialfn})). Here, we find that all components 
of the effective magnetic field are identically zero. Interestingly, the situation changes for interior Schwarzschild solutions, described by Eqs.(\ref{alphabeta}) and (\ref{extSch}). Here,
we find that the magnetic field at the interior is non-zero, 
and it does not fall off to zero at the matching radius, as one would have naively expected. In fact, at the matching radius $r=R$, we find that setting $c=1$,
\begin{equation}
b_2 = \frac{3R_s\epsilon}{8c^4R^2(R-R_s)}\sqrt{\epsilon^2 - c^4\left(1 - \frac{R_s}{R}\right)}
\end{equation}
This can be large for $R$ close to $R_s$ and for large values of $\epsilon$. Hence, there is a discontinuity 
for ${\vec b}$ at the matching radius. This is not surprising, since, as mentioned before, the fields depend on the derivatives of the metric, which can be discontinuous at the 
matching radius.  In fig.(\ref{SchIr}), we show the behavior of the field $b_2$ (in units of the observer's coordinates) as a function of the radial coordinate, for $\epsilon = 5$ and $6$ (in units of $c^2$). 
 The same qualitative feature is seen for a fermion in radial geodesic motion in a JNW background of Eq.(\ref{a1}). Here, we find that 
\begin{equation}
b_3 = -b_2 =\frac{\epsilon B^2 \left(\nu ^2-1\right) \sqrt{\epsilon ^2-c^4\left(1-\frac{B}{r}\right)^{\nu }}}{16 c^4 r^2(B-r)^2}
\end{equation}
Recalling that for radial geodesics in the JNW geometry, we have
\begin{equation}
{\dot t}= \frac{\epsilon}{c^2}\left(1-\frac{B}{r}\right)^{-\nu},~~~{\dot r} =\left[\frac{\epsilon^2}{c^2} - c^2\left(1-\frac{B}{r}\right)^{\nu}\right]^{\frac{1}{2}},
\end{equation}
we see that the field vanishes for ${\dot r}=0$ and in general can be large for large values of $\epsilon$ and small values of $r$.
 
Finally, we repeat the analysis for radial geodesics in BST backgrounds. First, let us recapitulate a few details \cite{dbs1}. For radial 
geodesics, it can be checked that 
\begin{equation}
{\dot t} = \frac{\epsilon}{c^2}\left(D + \frac{\alpha}{r}\right),~~~{\dot r} 
= \beta\left[\frac{\epsilon^2}{c^2} \left(D + \frac{\alpha}{r}\right) - c^2\right]^{1 \over 2}
\label{geot}
\end{equation}
Hence, the radial velocity is given by
\begin{equation}
v_{\rm rad} = \frac{\beta c \sqrt{r}}{\epsilon\left(\alpha + Dr\right)}
\left[\epsilon^2\left(Dr + \alpha\right) - c^4r\right]^{1 \over 2}
\label{geovrad}
\end{equation}
From Eq.(\ref{geovrad}) we see that the radial velocity thus becomes zero if the energy per unit mass satisfies 
\begin{eqnarray}
\epsilon^2 = \frac{c^4r}{\alpha + Dr}
\label{maxgammasq}
\end{eqnarray}
which can also be turned around to provide a maximum value of the radius at which the particle reaches zero velocity. 
Now, after transforming to Fermi normal coordinates, we calculate the components of the effective magnetic field and find that (with $b_0=b_1=0$),
\begin{equation}
b_2 = -b_3 = \frac{2\alpha\epsilon}{25c^4r^3}\sqrt{\frac{\epsilon^2\left(\alpha + Dr\right) - c^4r}{\alpha + Dr}}
\end{equation}
where we have set the parameter $\beta = 4/5$. Hence, the field is dependent on the energy of the observer, and 
is zero for the value of $\epsilon$ given in Eq.(\ref{maxgammasq}). Now if we use Eq.(\ref{bstalphad}) to obtain the field
in a galactic scenario, we obtain 
\begin{equation}
b_2 = \frac{GM\epsilon}{100c^2(v_c^{\rm max})^4r^3}\sqrt{\frac{\epsilon^2\left(GM + (v_c^{\rm max})^2r\right) - 8c^2(v_c^{\rm max})^4r}{GM + (v_c^{\rm max})^2r}}
\end{equation}
For a highly relativistic particle, this can be large. For example, if we take $M = 10^8M_{\odot}$, with $v_c^{\rm max} = 20{\rm Km/sec}$ and set $\epsilon = 10^{16}{\rm(m/sec)}^2$,
the we obtain, at a radius of $\sim 10^6{\rm m}$, $b_2 \sim 10^{-11}{\rm GeV}$. This can be significant in futuristic experiments.

\section{Discussions and Conclusions}

In this paper, we have considered the effective magnetic field due to curvature couplings in fermions. These arise purely due to gravity effects, and are different
in origin from intrinsic magnetic fields. Using Fermi normal coordinates,
we have computed these for the Schwarzschild, JNW and Bertrand space-times, for observers in circular and radial geodesics. Our results establish the qualitative
difference in the fields in various cases. We show that these fields can be large for specific parameter values in the theories, and hence might be indirectly observed in
futuristic experiments. 

For circular geodesics, whereas the direction of the magnetic field remains a constant as a function of the radial coordinate for the Schwarzschild and JNW backgrounds, this is not true 
for naked singularity backgrounds seeded by galactic dark matter. It was also shown that for Schwarzschild and JNW background, there is a large enhancement of the magnetic field 
not only near the photon sphere, but in regimes where circular orbits are stable. As we have mentioned, this might have implications for highly gravitating objects like black holes or neutron stars.
However, such enhancement is not observed in BST backgrounds. For the latter, large magnetic fields are seen by circular observers only close to the singularity. 

For observers on radial geodesics, our results show that whereas an external Schwarzschild observer will not see any magnetic field, the same is not true for internal Schwarzschild,
JNW, and BST observers. 
This is an important distinction between vacuum spaced-times and those seeded by matter, and should be investigated further. 
In these cases, the magnetic field is dependent on the observer's energy, and can be considerably large for highly relativistic observers. 

Before we conclude, we mention that our results for the JNW and the BST can be used to make a comparative analysis of gravitational effects on the Hydrogen atom spectra, that
has been well studied in the literature \cite{parker1}, \cite{parker2} by using degenerate perturbation theory. We will consider only non relativistic energy shifts and will simply state 
our main results here. For the $1S$ and $2S$ states of the Hydrogen atom, in Schwarzschild backgrounds, the non relativistic energy shifts are zero (Eq.(6.1) and (6.15) of 
\cite{parker3}). The essential reason is that these are proportional to the $R_{00}$ component of the Ricci tensor, which is identically zero for radial and circular geodesics 
in the Schwarzschild geometry, in Fermi normal coordinates. For JNW and BST backgrounds, this ceases to be the case (see Appendix B for the explicit expression of
$R_{00}$ for circular geodesics in BST geometries). Hence, for these naked singularity backgrounds, the $1S$ and $2S$ levels of Hydrogen will receive gravitational corrections, contrary
to the situation for black hole backgrounds. 

We have calculated this shift and find that for radial geodesics, it can be large near the singularity for JNW space-times and for small radial distances for BSTs. We also 
calculated the shifts in the energy of some of the other states listed in \cite{parker3} (Eqs.(6.1) to (6.23) of that paper) and find that in general, gravitational effects in JNW 
and BST backgrounds will be large only near the naked singularity, for radial geodesics. 

We also calculated the energy shifts for circular geodesics, and find that for JNW backgrounds, shifts in the Hydrogen atom spectrum are qualitatively similar to those
in Schwarzschild backgrounds, i.e they become large only near the photon sphere (a result that is evident for the Schwarzschild geometry from Eqs.(6.15) - (6.23) of 
\cite{parker3}) or near the naked singularity if the photon sphere is absent. The second situation is more interesting as it allows for stable orbits, as we have
seen in section 2.4. For BSTs, these are large only near the central singularity. 

To conclude, we have presented a comprehensive GR analysis of effective magnetic fields seen by fermions in geodesic motion in curved space-times. Our results complement and add to 
the existing literature on the subject. We show evidence of large magnetic fields induced by gravity near a very massive gravitating object or near the centre of a galaxy. We have also 
demonstrated qualitative differences in magnetic field, depending on the nature of space-time. Experimental signatures of these effects in futuristic experiments may be important to analyze. 

\begin{center}
{\bf Acknowledgements}
\end{center}
It is a pleasure to thank Kaushik Bhattacharya for very helpful discussions.

\appendix
\section{Appendix A}

The non-zero components of Riemann tensor in Fermi normal coordinates for circular geodesics in the background of Schwarzschild black hole 
(at $\theta=\pi/2$) are listed below. These also appear in \cite{collas}.
\begin{eqnarray}
 R_{0101} &=& {r+3(r-R_S)\cos(2\delta \phi)\over 2 r^3 (2r-3R_S)}R_S\nonumber\\
 R_{0103} &=& {3 R_S(r-R_S)\over 2 r^3 (2r-3R_S)}\sin(2\delta \phi)\nonumber\\
 R_{0113} &=& -{3 R_S \sqrt{R_S(r-R_S)}\over \sqrt{2} r^3(2r-3R_S)}\cos(\delta
\phi)\nonumber\\
 R_{0202} &=& -\frac{R_S}{r^2(2r-3R_S)}\nonumber\\
 R_{0212} &=& {3 R_S \sqrt{R_S(r-R_S)}\over \sqrt{2} r^3(2r-3R_S)}\sin(\delta
\phi)\nonumber\\
 R_{0303} &=& {r-3(r-R_S)\cos(2\delta \phi)\over 2 r^3 (2r-3R_S)}R_S\nonumber\\
  R_{0223} &=& -R_{0113}\nonumber\\
 R_{0313} &=& -R_{0212}\nonumber\\
 R_{1212} &=& -R_{0303}\nonumber\\
 R_{1223} &=& -R_{0103}\nonumber\\
 R_{1313} &=& -R_{0202}\nonumber\\
 R_{2323} &=& -R_{0101}\nonumber
\end{eqnarray}
where, \hspace{2mm} $R_S={2GM\over c^2}$, \hspace{1mm} and \hspace{1mm} $\delta =
\sqrt{1-{3R_S\over 2r}}=\sqrt{1-{3GM\over c^2r}}$. \\
In Fermi normal coordinates, it is easy to check that $R_{00}=0$.

\section{Appendix B}

The non-zero components of Riemann tensor in Fermi normal coordinates for circular geodesics in the background of Bertrand space-time 
(at $\theta=\pi/2$) are listed below:
\begin{eqnarray}
 R_{0101} &=& \alpha  \beta ^2\frac{D r+ (\alpha +3 D r)\cos 2 \phi}{2 r^2(\alpha +D
r) (\alpha +2 D r)}\nonumber\\
 R_{0103} &=& \alpha  \beta ^2\frac{(\alpha +3 D r)\sin 2\phi}{2r^2(\alpha +D r)
(\alpha +2 D r)}   \nonumber\\
 R_{0113} &=& -\frac{\alpha ^{3/2} \beta ^2 (\alpha +4 D r)\cos \phi }{r^2
\Big(2(\alpha +D r)\Big)^{3/2} (\alpha +2 D r)}\nonumber\\
 R_{0202} &=& -\frac{\alpha }{r^2 (\alpha +2 D r)}\nonumber\\
 R_{0212} &=& \frac{\sqrt{\alpha } \Big(\alpha  \left(2-\beta^2\right)
 +2 D r \left(1-\beta ^2\right)\Big)\sin \phi }{r^2\sqrt{2(\alpha +D r)} (\alpha +2
D r)}\nonumber\\
 R_{0223} &=& \frac{\sqrt{\alpha } \Big(\alpha  \left(2-\beta^2\right)
 +2 D r \left(1-\beta ^2\right)\Big)\cos \phi }{r^2\sqrt{2(\alpha +D r)} (\alpha +2
D r)}\nonumber\\
 R_{0303} &=& \alpha  \beta ^2 \frac{D r-(\alpha +3 D r) \cos 2 \phi}{2
   r^2 (\alpha +D r) (\alpha +2 D r)}\nonumber\\
 R_{0313} &=& -\frac{\alpha ^{3/2} \beta ^2  (\alpha +4 D r) \sin \phi}{r^2
\Big(2(\alpha +D r)\Big)^{3/2} (\alpha +2 D r)}\nonumber\\
 R_{1212} &=& \frac{\beta ^2 (\alpha +2 D r) (3 \alpha+2 D r)-4 (\alpha +D r)^2}{2
r^2 (\alpha +D r)
 (\alpha +2 Dr)}\sin ^2\phi \nonumber\\
 R_{1223} &=& \frac{\beta ^2 (\alpha +2 D r) (3 \alpha+2 D r)-4 (\alpha +D r)^2}{4
r^2 (\alpha +D r)
 (\alpha +2 Dr)}\sin 2 \phi  \nonumber\\
 R_{1313} &=& \frac{\alpha ^2 \beta ^2 (\alpha +4 D r)}{4 r^2 (\alpha +D r)^2(\alpha
+2 D r)}\nonumber\\
 R_{2323} &=& \frac{\beta ^2 (\alpha +2 D r) (3 \alpha+2 D r)-4 (\alpha +D r)^2}{2
r^2 (\alpha +D r)
 (\alpha +2 Dr)}\cos ^2\phi \nonumber
\end{eqnarray}
For this space-time, it can be checked that in Fermi normal coordinates,
\begin{eqnarray}
 R_{00} = -\frac{\alpha  \Big(\alpha +Dr \left(1-\beta ^2 \right)\Big)}{r^2
   (\alpha +D r) (\alpha +2 D r)}\nonumber
\end{eqnarray}

\end{document}